\documentclass[12pt]{iopart}

\usepackage{iopams}
\usepackage{bm} 					
\usepackage{upgreek}				
\usepackage{dsfont}
\usepackage{graphicx} 				
	\usepackage{color}				
	\usepackage{transparent}		
\usepackage{fancyhdr}				

\usepackage{setspace}

\usepackage{dsfont}				
\usepackage{microtype}		
\usepackage{placeins}			
\usepackage{acronym}			

\let\SQRT\sqrt
\renewcommand{\sqrt}[1]{\ensuremath{\SQRT{#1} \;}}

\let\REALPART\Re
\renewcommand{\Re}[1]{\ensuremath{\REALPART\left\{ #1 \right\} }}
\let\IMAGPART\Im
\renewcommand{\Im}[1]{\ensuremath{\IMAGPART\left\{ #1 \right\} }}

\newcommand{\eqref}[1]{(\ref{#1})}

\catcode`_=\active
\newcommand_[1]{\ensuremath{\sb{\mathrm{#1}}}}

\newcommand{\intfin}[3]{\ensuremath{\int\sb{#2}^{#3}} \textup{d}#1 \;} 
\newcommand{\intinf}[1]{\ensuremath{\int\sb{-\infty}^{\infty}} \textup{d}#1 \;} 

\newcommand{\mod}[1]{\ensuremath{\left| #1 \right|}} 
\newcommand{\td}{\ensuremath{\left(t\right)}} 
\newcommand{\trans}[1]{\ensuremath{ \left.#1\right.^{\mathrm{T}}}} 

\newcommand{\identity}{\ensuremath{ \mathds{1} }} 

\newcommand{\Br}[1]{\ensuremath{\left( #1 \right)}} 
\newcommand{\Sq}[1]{\ensuremath{\left[ #1 \right]}} 
\newcommand{\Cu}[1]{\ensuremath{\left\{ #1 \right\}}} 

\newcommand{\ket}[1]{\ensuremath{ \left|  #1 \right> }}         
\newcommand{\expect}[1]{\ensuremath{\left< #1 \right>}}
\newcommand{\comm}[2]{\ensuremath{\left[ #1,#2 \right]}}
\newcommand{\n}{\ensuremath{\bar{n}}} 

\newcommand{\I}{\mathrm{i}} 					
\renewcommand{\e}{\mathrm{e}} 					

\newcommand{\half}{\ensuremath{\frac{1}{2}}}   

\newcommand{\Schrod}{Schr\"{o}dinger} 

\newcommand{\via}{\textit{via}} 
\newcommand{\ie}{\textit{i.e.}} 
\newcommand{\eg}{\textit{e.g.}} 
\newcommand{\viz}{\textit{viz}.} 
\newcommand{\cf}{\textit{cf}.} 
\newcommand{\ala}{\textit{\`{a} la}} 
\renewcommand{\etal}{\textit{et al.}} 

\newcommand{\naive}{na\"\i{}ve} 

\begin{document}

\title{Rapid mechanical squeezing with pulsed optomechanics}
\author{James S. Bennett, and Warwick P. Bowen}
\address{Australian Research Council Centre of Excellence for Engineered Quantum Systems (EQuS), The University of Queensland, St Lucia, QLD 4072, Australia}
\ead{james.bennett2@uqconnect.edu.au}

\begin{abstract}
Macroscopic mechanical oscillators can be prepared in quantum states and coherently manipulated using the optomechanical interaction. This has recently been used to prepare squeezed mechanical states. However, the scheme used in these experiments relies on slow, dissipative evolution that destroys the system's memory of its initial state. In this paper we propose a protocol based on a sequence of four pulsed optomechanical interactions. In addition to being coherent, our scheme executes in a time much shorter than a mechanical period. We analyse applications in impulsive force sensing and preservation of \Schrod{} cat states, which are useful in continuous-variable quantum information protocols.
\end{abstract}

\date{\today}

\maketitle

\section{Introduction} \label{Sec:Introduction}

Recent experiments in optomechanics have undeniably demonstrated that macroscopic mechanical resonators are quantum systems. Achievements such as near-ground-state cooling and controllable single-phonon excitation \cite{OConnell2010, Chu2017}, electromechanical state writing and retrieval \cite{Palomaki2013}, and entanglement of microwave and mechanical modes \cite{Palomaki2013a} show the exciting potential of mechanical oscillators for both fundamental studies of physics \cite{Chen2013,Treutlein2014a} and in future quantum technologies \cite{Kimble2008,Tsukanov2011,Muschik2011,Stannigel2012,Schmidt2012}. Mechanical squeezing---the reduction of a motional quadrature's variance below that of the ground state---is an important quantum resource for these applications \cite{Braunstein2005}.

Approaches to mechanical squeezing generally use parametric amplification \cite{Agarwal1991,Szorkovszky2011,Liao2011a,Agarwal2016}, which takes many mechanical periods to generate squeezing; non-unitary processes such as measurement \cite{DiFilippo1992,Vitali2002,Ruskov2005,Filip2005,Vanner2011,Vanner2013}; or lossy evolution towards a squeezed steady-state (as proposed by \cite{Agarwal2016,Clerk2008b,Jahne2009,Mari2009,Nunnenkamp2010,Wang2016}). Some near-unitary squeezing schemes have been proposed, but they are best suited to trapped ions, neutral atoms, levitated optomechanical systems \cite{Bhattacharya2017}, and other platforms where the mechanical resonance frequency can be rapidly tuned over a range comparable to or larger than itself \cite{Janszky1992}.

The first experimental demonstration of mechanical squeezing to below the ground-state variance was performed by Meekhof \etal{} \cite{Meekhof1996}, who made use of a parametric amplification scheme to squeeze the motion of a trapped ion. More recently, solid-state oscillators have been squeezed by Wollman \etal{} \cite{Wollman2015} and Pirkkalainen \etal{} \cite{Pirkkalainen2015a} using a two-tone optomechanical driving scheme that permits a squeezed steady state. As already noted, these operations are much slower than the mechanical period.

In this paper we describe a squeezing operation created by a series of four pulsed optomechanical interactions. Our scheme can squeeze an arbitrary mechanical state extremely quickly---coherently generating squeezing over timescales much shorter than the mechanical period. In contrast to schemes based on rapidly tuning the mechanical resonance, it also operates at a fixed mechanical frequency, making it suitable for use with well-developed solid-state optomechanical devices.

In section~\ref{Sec:PulsedQND} we outline how pulsed optomechanics can be used to generate a quantum non-demolition (QND) interaction. We then use this to construct our squeezer in section~\ref{Sec:Construct}. The performance of the squeezer for Gaussian and non-Gaussian inputs is discussed in sections \ref{Sec:Gaussian} and \ref{Sec:NonGaussian} respectively. Finally, we provide some simple examples of how this rapid squeezer can be used, either to improve the sensing of impulsive forces (section~\ref{Sec:Impulse}), or to aid the storage of mechanical \Schrod{} cat states (section~\ref{Sec:Cats}).

\section{Pulsed optomechanical quantum nondemolition interactions} \label{Sec:PulsedQND}

An optomechanical system is composed of photonic and phononic resonators coupled such that motion of the mechanical element modulates the resonance frequency of the optical cavity \cite{BowenMilburn}. We describe the mechanical oscillator using its dimensionless position ($X_{M}$) and momentum ($P_{M}$) operators, which obey the commutation relation $\comm{X_{M}}{P_{M}} = 2\I$ (\ie{} `$\hbar = 2$ units'). Its angular frequency is $\omega_{M}$. The optical mode can be described by analogous operators that represent the amplitude ($x_{L}$) and phase ($p_{L}$) fluctuations of the field.

If the illumination consists of short pulses it is convenient to construct averaged quadrature operators $X_{L}$ and $P_{L}$ from the input and output optical fields. The instantaneous input and output fluctuation operators are connected to the intracavity field by the input--output relation $x_{L}^{\mathrm{out}} = \sqrt{\kappa} x_{L}-x_{L}^{\mathrm{in}}$ \cite{WallsMilburn}, where $\kappa$ is the cavity's energy decay rate. Weighting these operators by the normalised envelope of the pulse $f\td$, \viz{}
\begin{eqnarray}
		X_{L}^{\mathrm{in} \; \Br{\mathrm{out}}} & = & \intinf{t} \mod{f\td}x_{L}^{\mathrm{in} \; \Br{\mathrm{out}}}\td \label{Eqn:PulseX} \\
		P_{L}^{\mathrm{in} \; \Br{\mathrm{out}}} & = & \intinf{t} \mod{f\td}p_{L}^{\mathrm{in} \; \Br{\mathrm{out}}}\td \label{Eqn:PulseP},
\end{eqnarray}
yields pairs of quadratures $\comm{X_{L}^{\mathrm{in}}}{P_{L}^{\mathrm{in}}} = \comm{X_{L}^{\mathrm{out}}}{P_{L}^{\mathrm{out}}}= 2\I$.

Letting the duration of the pulse be extremely short compared to the mechanical period $2\pi/\omega_{M}$ and the typical thermalisation time (\cf{} section~\ref{Sec:LossyPerformance}) allows us to neglect the damped mechanical evolution over the duration of the pulse. The linearised optomechanical interaction then becomes \cite{Vanner2011}
\begin{eqnarray}
		X_{M}^{\prime} =  X_{M}, & \;\;\; & P_{M}^{\prime} = P_{M} + \chi X_{L} \label{Eqn:QNDX} \\
		X_{L}^{\prime} =  X_{L}, & \;\;\; & P_{L}^{\prime} = P_{L} + \chi X_{M}. \label{Eqn:QNDP}
\end{eqnarray}
We have marked output operators with primes, and introduced the interaction strength $\chi = -8g_{0}\sqrt{N}/\kappa$. Here $g_{0}$ is the single-photon optomechanical coupling rate \cite{BowenMilburn}, $N$ is the mean photon number in the pulse envelope, and $\kappa$ is the linewidth of the optical mode. This result holds under three approximations. Firstly, we require that the optomechanical interaction with a single photon is weak, such that $g_{0} \ll \omega_{M}$. Secondly, the optical linewidth must be sufficient to reach the deep unresolved sideband regime, $\omega_{M} \ll \kappa$. This ensures that the intracavity photon number is high during the pulse, and that the cavity does not distort it. Finally, we assume that the optomechanical coupling targets a single mechanical mode. Discussion of this approximation is given in \ref{App:AdditionalMechModes}. We have also omitted a `classical' momentum displacement acquired by the mechanical oscillator, which can be subtracted from the interaction by including an open-loop displacement operation.

Equations~\eqref{Eqn:QNDX} and \eqref{Eqn:QNDP} are said to be an $X$--$X$ quantum non-demolition (QND) interaction because they leave both $X$ operators unchanged; the operation imprints information only on the $P$ operators. This will be the basis of our proposal. As such, this scheme could be adapted to any system permitting such a QND interaction.

\section{Constructing a squeezer} \label{Sec:Construct}

For the moment, let us imagine that in addition to the $X$--$X$ interaction described above we also have access to one which preserves both momenta (a $P$--$P$ interaction). The latter does not naturally arise in pulsed optomechanics, but we will show how to approximate it in section~\ref{Sec:PP}. It will be convenient to write these interactions as a homogeneous linear equation acting on the `vector' of operators $\bm{X} = \trans{\Br{X_{M},P_{M},X_{L},P_{L}}}$, \viz{}
\begin{eqnarray}
	\bm{X}^{\prime} & = & \Br{\begin{array}{cccc}
	1 & 0 & 0 & 0 \\
	0 & 1 & \chi & 0 \\
	0 & 0 & 1 & 0 \\
	\chi & 0 & 0 & 1 \end{array}}\bm{X} \;\;\; \Br{X\mathrm{-}X\; \mathrm{QND}}, \\
	\bm{X}^{\prime} & = & \Br{\begin{array}{cccc}
	1 & 0 & 0 & \chi \\
	0 & 1 & 0 & 0 \\
	0 & \chi & 1 & 0 \\
	0 & 0 & 0 & 1 \end{array}}\bm{X} \;\;\; \Br{P\mathrm{-}P\; \mathrm{QND}},
\end{eqnarray}
where the output vectors bear a prime. These matrices will be denoted by $\mathcal{M}_{XX}\Br{\chi}$ and $\mathcal{M}_{PP}\Br{\chi}$ respectively.

Consider the sequence of three QND interactions defined by
\begin{equation}
	\mathcal{O} = \mathcal{M}_{XX}\Br{\chi_{3}}\mathcal{M}_{PP}\Br{\chi_{2}}\mathcal{M}_{XX}\Br{\chi_{1}}.
	\label{Eqn:O}
\end{equation}	
$\mathcal{O}$ has diagonal blocks given by
\[
	\mathcal{D} = \Br{\begin{array}{cc}
	1+\chi_{1}\chi_{2} & 0 \\
	0 & 1+\chi_{2}\chi_{3}
	\end{array}},
\]
and antidiagonal blocks
\begin{equation}
	\mathcal{A} = \Br{\begin{array}{cc}
	0 & \chi_{2} \\
	\chi_{1}+\chi_{3}\Br{1+\chi_{1}\chi_{2}} & 0
	\end{array}}. \label{Eqn:Antidiagonal}
\end{equation}
Selecting the interaction strengths such that $\chi_{2} = -\Br{\chi_{1}^{-1}+\chi_{3}^{-1}}$ has two effects; it ensures that $\mathcal{D}$ is brought into a diagonal form with unit determinant, and also cancels the lower left element of $\mathcal{A}$. Thus the transformation is (here showing only the mechanical output)
\begin{eqnarray}
		X_{M}^{\prime} & = & -\frac{\chi_{1}}{\chi_{3}} X_{M} -\Br{\chi_{1}^{-1}+\chi_{3}^{-1}} P_{L}, \label{Eqn:IdealXTrans}\\
		P_{M}^{\prime} & = & -\frac{\chi_{3}}{\chi_{1}} P_{M}. \label{Eqn:IdealPTrans}
\end{eqnarray}
If the optical ancilla is squeezed sufficiently strongly that the added noise is well below the vacuum (\ie{} variance of $P_{L}$ satisfies $V_{sq} \ll \Sq{\chi_{1}\chi_{3}/\Br{\chi_{1}+\chi_{3}}}^{2}$) this becomes an effectively unitary squeezing operation. It position squeezes if $\mod{\chi_{1}} < \mod{\chi_{3}}$, and momentum squeezes when $\mod{\chi_{1}} > \mod{\chi_{3}}$. Many readers will recognise the similarity between this concept and the feedforward-based optical squeezing scheme discovered by Filip, Marek, and Andersen \cite{Filip2005}.

It is also possible to draw parallels with previous work on pulsed optomechanics \cite{Bennett2016} that relied on essentially the same QND sequence as used here; to see this, note that setting $\chi_{1} = \chi_{2} = \chi_{3} = -1$ in equation~\eqref{Eqn:O} yields the same state-swap interaction as in \cite{Bennett2016}. The methods used in this paper could therefore also be applied to speed up the state-swap procedure presented in \cite{Bennett2016}.

\subsection{Approximating the $P$--$P$ interaction} \label{Sec:PP}

As stated above, the pulsed optomechanics toolbox unfortunately does not contain a $P$--$P$ QND interaction. Evidently we will have to content ourselves with an approximation. To obtain one we will first note that a $P$--$P$ interaction is related to a $P$--$X$ interaction (one that preserves $P_{M}$ and $X_{L}$) $\mathcal{M}_{PX}$ \via{} rotations of the optical mode, $R_{L}$, \viz{}
\[
	\mathcal{M}_{PP} = R_{L}\Br{\pi/2}\mathcal{M}_{PX}R_{L}^{-1}\Br{\pi/2}.
\]
Note that the rotations of the optical noise can, in principle, be performed arbitrarily quickly. This is done by displacing the optical pulse \eg{} by interfering it with a bright coherent pulse on a highly asymmetric beamsplitter \cite{Furusawa1998}.

At this stage we may employ the work of Khosla \etal{} \cite{Khosla2017} to obtain an approximate $\mathcal{M}_{PX}$ transformation by combining two $X$--$X$ QND interactions and a short mechanical rotation $R_{M}\Br{\varphi}$. Optimising the strength of the latter pulse with respect to the former, $\lambda$, yields 
\begin{eqnarray}
	\overline{\mathcal{M}}_{PX} & = & \mathcal{M}_{XX}\Br{-\frac{\lambda}{\cos\varphi}} R_{M}\Br{\varphi} \mathcal{M}_{XX}\Br{\lambda} \label{Eqn:MPX}\\
	& = & R_{M}\Br{\varphi} \Br{\begin{array}{cccc}
					1 & 0 & \lambda\tan\varphi & 0 \\
					 0 & 1 & 0 & 0 \\
					 0 & 0 & 1 & 0 \\
					 0 & -\lambda\tan\varphi & -\lambda^{2}\tan{\varphi} & 1
				\end{array}}, \nonumber
\end{eqnarray}
where we have used an over-line on the matrix to indicate that it is an approximation of the desired interaction. Up to a mechanical rotation, this has the form of a $P$--$X$ QND interaction with strength $\chi_{2} = +\lambda\tan\varphi$, followed\footnote{Or indeed preceded by it, because the Hamiltonians commute.} by a Kerr interaction with a Hamiltonian of the form $X_{L}^{2}$.

We might expect that we can use this to achieve an approximate squeezer $\overline{\mathcal{O}}$ by making the replacement $\mathcal{M}_{PP} \rightarrow \overline{\mathcal{M}}_{PP}$ in equation~\eqref{Eqn:O}, with
\[
	\overline{\mathcal{M}}_{PP}\Br{\chi_{2}} = R^{-1}_{L}\Br{\pi/2}\overline{\mathcal{M}}_{PX}\Br{\chi_{2}}R_{L}\Br{\pi/2}.
\]
However, this choice results in the Kerr nonlinearity becoming mapped onto the output mechanical state through the final $X$--$X$ interaction. To avoid this undesired nonlinearity we can include an additional optical rotation that rotates the effect of the Kerr interaction onto a quadrature that does not participate in the final QND interaction. The necessary rotation angle $\theta$ obeys $\tan\theta = -\lambda^{2}\tan\varphi$.

\begin{figure}
\centering
\def\svgwidth{0.9\columnwidth}
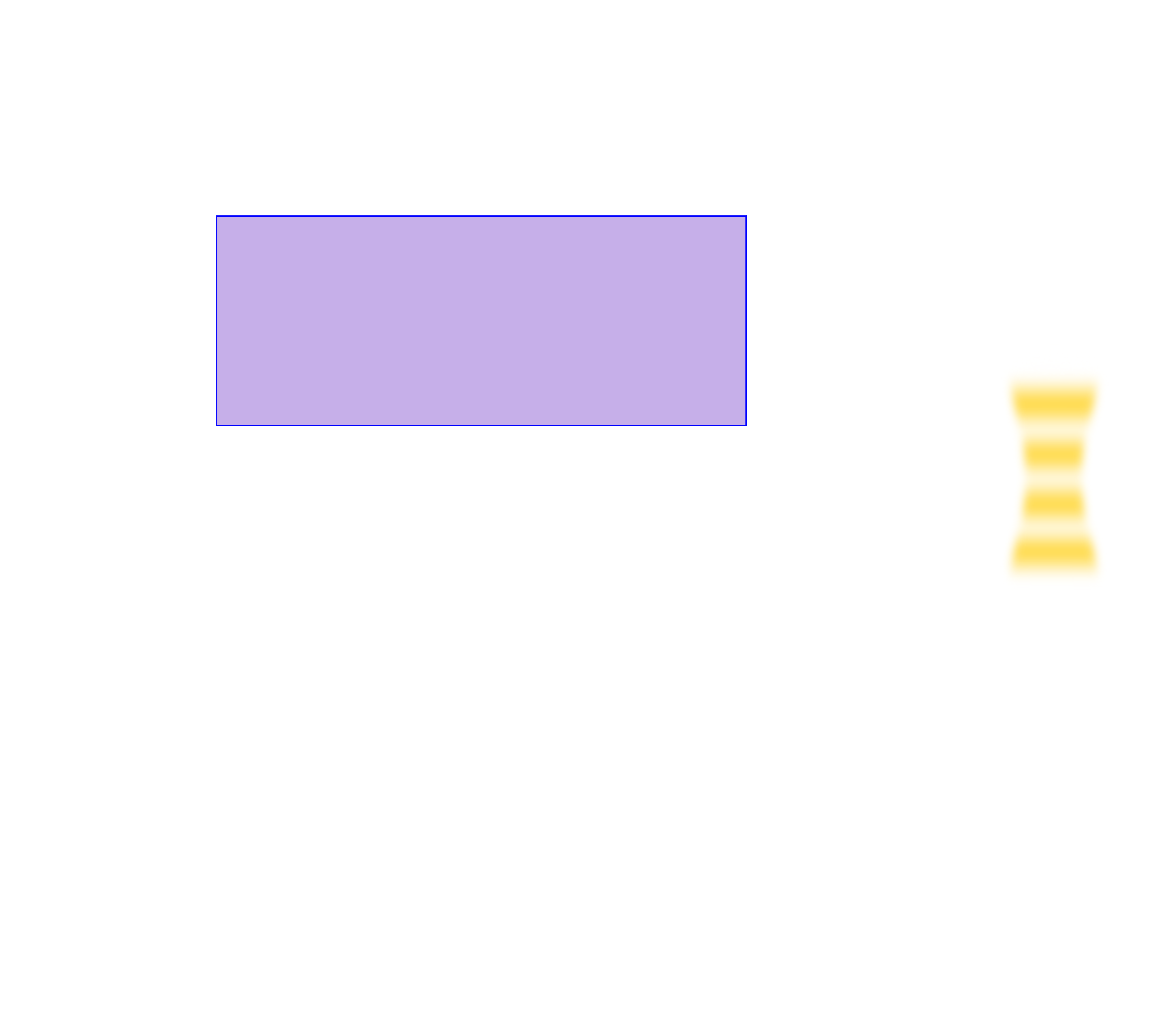
\caption{\label{Fig:Scheme}
\textit{a})~Schematic of idealised fast squeezer based on consecutive QND interactions. Time flows from left to right. The top rail represents the mechanical mode, and the bottom is a squeezed optical ancilla. The first abstraction step (purple, \textit{i}) is to realise the $P$--$P$ interaction using optical rotations and a $P$--$X$ interaction. The yellow panel (\textit{ii}) shows how the latter can be approximated with the $X$--$X$ QNDs provided by optomechanics, and a mechanical rotation through an angle $\varphi$.\\
\textit{b})~A potential architecture for realisation of an optomechanical fast squeezer. We assume that the optical cavities support identical, polarisation-degenerate modes, and that the oscillator is perfectly reflective. Grey lines indicate the beam path; all beamsplitters are polarising (transmit the input polarisation); partially silvered mirrors (dotted) are used to achieve optical rotations (using control pulses, CP); and all waveplates (purple rectangles) are $\lambda/4$ plates. The input squeezed pulse (left) interacts with both cavities before being shunted into the delay line. The pulse then back-tracks from upper to lower cavities, and is finally outcoupled through the input port.\\
\textit{c})~Schematic of lossy protocol corresponding to a realistic experiment \ie{} the apparatus in panel~\textit{b}). Mechanical loss is introduced by a hot bath during the delay time $t = \varphi/\sigma\omega_{M}$: the input thermal noise is characterised by the bath occupancy $\n_{M}$. We model the optical loss as a beamsplitter with reflectivity $\epsilon$ situated in the delay line. This beamsplitter couples in thermal noise with an occupancy of $\n_{L}$.
}
\end{figure}

Combining all of these yields
\begin{eqnarray}
	\overline{\mathcal{O}} & = & \mathcal{M}_{XX}\Br{\chi_{3}}R_{L}\Br{\theta}\overline{\mathcal{M}}_{PP}\Br{\chi_{2}}\mathcal{M}_{XX}\Br{\chi_{1}}.
	\label{Eqn:Obar} 
\end{eqnarray}
This full scheme is depicted in Fig.~\ref{Fig:Scheme},~panel~\textit{a}).

Selecting
\[
	\chi_{3} = -\frac{\chi_{1}\sqrt{1+\lambda^{4}\tan^{2}\varphi}}{\cos\varphi + \lambda\chi_{1}\sin\varphi}
\] 
brings the diagonal components of $\overline{\mathcal{O}}$ into the desired form, in addition to cancelling off some unwanted contributions to $P_{M}^{\prime}$. This leaves the net mechanical evolution (not including a rotation through $\varphi$, for clarity)
\begin{eqnarray}
		X_{M}^{\prime} & = & \Br{1+\lambda\chi_{1}\tan\varphi} X_{M} + \frac{\lambda\chi_{1}\tan^{2}\varphi}{1+\lambda\chi_{1}\tan\varphi} P_{M} \label{Eqn:ApproxXTrans}\\
		& & {}+\Cu{\chi_{1}\tan\varphi X_{L} + \lambda\tan\varphi P_{L}}, \nonumber\\
		P_{M}^{\prime} & = & \frac{1}{1+\lambda\chi_{1}\tan\varphi} P_{M}. \label{Eqn:ApproxPTrans}
\end{eqnarray}
We will refer to this as the `idealised' QND squeezer---it is close to, but not equal to, a true unitary squeezer.

\subsection{Idealised squeezer} \label{Sec:IdealBreakdown}

The diagonal elements of equations~\eqref{Eqn:ApproxXTrans} and \eqref{Eqn:ApproxPTrans} correspond to rescaling the momentum by a factor $\mu = \Br{1 + \lambda\chi_{1}\tan\varphi}^{-1}$, with a corresponding scaling of $X$ by $\mu^{-1}$. This is exactly the unitary squeezer we desired.

In addition, equation~\eqref{Eqn:ApproxXTrans} (for the mechanical position) contains two undesirable terms. The first ($\Br{1-\mu}\tan\varphi P_{M}$) arises from the fact that we are unable to achieve a true $P$--$P$ QND interaction using optomechanics. The second corresponds to noise originating from the optical ancilla, as in equation~\eqref{Eqn:IdealXTrans}.

We can succinctly write the optical noise term as $AX_{L}^{\Br{\phi}}$, where the quadrature angle is $\phi = \arctan\Br{\lambda/\chi_{1}}$ and the coefficient is $A = \sqrt{\lambda^{2}+\chi_{1}^{2}} \tan\varphi $. Minimising $A$ at fixed $\mu$ and $\varphi$ (\ie{} fixed $\chi_{1}\lambda$ product) can be achieved by selecting $\lambda = \pm\chi_{1}$, with like signs ($\lambda=\chi_{1}$) resulting in momentum squeezing ($\mu < 1$) and unlike signs ($\lambda=-\chi_{1}$) in position squeezing ($\mu > 1$). The optimum ancilla squeezing angles are $\phi_{opt} = \pi/4$ and $-\pi/4$ respectively. Using this we can compactly write $\phi_{opt} = {\mathrm{sgn}\Cu{1-\mu}\frac{\pi}{4}}$, and thus
\begin{eqnarray}
		X_{M}^{\prime} & = & X_{M}/\mu + \Br{1-\mu}\tan\varphi P_{M} + \sqrt{2\mod{1-\mu^{-1}}\tan\varphi} X_{L}^{\Br{\phi_{opt}}},  \label{Eqn:XOpt}\\
	P_{M}^{\prime} & = & \mu P_{M}. \label{Eqn:POpt}
\end{eqnarray}
In doing this we have made use of the sign function ($\mathrm{sgn}$) and tacitly restricted ourselves to positive values of $\mu$. Note that this operation becomes a perfect squeezer in the $\varphi \rightarrow 0$ limit, but maintaining $\mu$ requires a corresponding increase in the pulse strengths. This will be discussed further in section~\ref{Sec:PhotonNumber}. For finite $\varphi$ the performance of the squeezer (in the absence of loss) is limited by optical squeezing and by the initial mechanical (momentum) noise. The former dominates in the $\mu > 1$ regime, and the latter when $\mu < 1$.
For finite $\varphi$ the performance of the squeezer (in the absence of loss) is typically limited by optical squeezing---particularly for $\mu > 1$---and by the initial mechanical (momentum) noise---especially for $\mu < 1$.
 These asymmetries will become quite apparent in section~\ref{Sec:LosslessPerformance} when we calculate the squeezer's fidelity.

\section{Performance: Gaussian inputs} \label{Sec:Gaussian}

To begin with, let us restrict ourselves to considering only Gaussian input states. The ideal (lossless) case is explored in section~\ref{Sec:LosslessPerformance}, then mechanical and optical damping are introduced in section~\ref{Sec:LossyPerformance}.

\subsection{Lossless performance} \label{Sec:LosslessPerformance}

To quantify the performance of our squeezer we will calculate the fidelity $\mathcal{F}$ of its output states \ie{} the overlap between the actual output state ($V^{\prime}$) and the desired output ($V_{tar}$). For zero-mean states this is given by \cite{Weedbrook2012}
\[
	\mathcal{F} = 2\Br{
		\begin{array}{c}
			\sqrt{\mod{V^{\prime} + V_{tar}} + \Br{\mod{V^{\prime}}-1}\Br{\mod{V_{tar}}-1}} \\
			-\sqrt{\Br{\mod{V^{\prime}}-1}\Br{\mod{V_{tar}}-1}}
			\end{array}}^{-1}.
\]
If the input state is pure ($\mod{V}_{tar} = 1$) this reduces to
\begin{equation}
	\mathcal{F}_{pure} = \Br{1+\mu \mod{1-\mu} V_{p} \Sq{V_{sq} + \frac{\mu\mod{1-\mu}}{4}V_{p}\tan\varphi}\tan\varphi}^{-1/2}
	\label{Eqn:PureFidelity}
\end{equation}
where the initial state's momentum variance is $V_{p}$, and $V_{sq}$ is the squeezed variance of the optical input. This result will allow us to determine whether our scheme can truly perform a coherent squeezing operation.

Imagine the following process that aims to mimic squeezing using measurement and feedforward (\ie{} an incoherent process). We first clone the mechanical input state, then perform orthogonal homodyne measurements on each copy. The measurement results are then fed forward---with appropriate gain---onto an ancillary state. A phase-independent cloner is the most appropriate choice because, in general, the input state is unknown. By using the optimal phase-independent Gaussian cloner of \cite{BraunsteinPati}, taking the ancilla to be coherent (\ie{} `classical'), and employing the Brunn--Minkowski inequality (for $2\times 2$ matrices $A$ and $B$) \cite{Marcus1968}
\[
	\sqrt{\mod{A+B}} \geq \sqrt{\mod{A}} + \sqrt{\mod{B}}
\]
we obtain the state-independent bound 
\begin{equation}
	\mathcal{F}_{class}\Br{\mu} \leq \Br{1+\sqrt{\half+\frac{\mu^{2}+\mu^{-2}}{4}}}^{-1}
	\label{Eqn:ClassicalLimit}
\end{equation}
on the output fidelity. We will call this the `classical' fidelity. The largest possible value of $\mathcal{F}_{class}$ is one half. Thus we may conclude that $\mathcal{F}_{pure} > 1/2$ is a \textit{sufficient} condition for the squeezer to have operated with some degree of quantum coherence.

As we see in Fig.~\ref{Fig:Fidelity}, it is certainly possible to achieve $\mathcal{F} > \mathcal{F}_{class}$, even in the presence of loss (discussed in section~\ref{Sec:LossyPerformance}). Equal distances along the $\mu$ axis correspond to equal changes in the variance of the antisqueezed quadrature \eg{} for a perfect squeezer $\mu = 2$~($\log\mu \approx 0.7$) and $\mu = 1/2$~($\log\mu \approx -0.7$) produce equal amounts of squeezing along orthogonal axes. $\mu = 1$ corresponds to doing no squeezing, which is why the infidelity drops to zero at this point (in the absence of loss). The asymmetry between $\mu \rightarrow 0$ and $\mu \rightarrow \infty $ is due to the contribution of the initial momentum fluctuations, as previously noted.

\subsection{Lossy performance} \label{Sec:LossyPerformance}

In any realistic situation we must consider irreversible loss on both the optical and mechanical modes. There are many potential points within the protocol at which these could be introduced; we will choose them by considering the apparatus depicted in Fig.~\ref{Fig:Scheme},~panel~\textit{b}).

The unavoidable interaction between the mechanical oscillator and its thermal bath is the most obvious source of decoherence. In this paper we are primarily interested in evolution times that are short compared to the mechanical period, so we must use the momentum-damped Langevin equations
\begin{eqnarray}
	\dot{X}_{M} & = & \omega_{M}P_{M}, \label{Eqn:Xdot} \\ 
	\dot{P}_{M} & = & -\omega_{M}X_{M}-\Gamma P_{M} + \sqrt{2\Gamma}\xi\td, \label{Eqn:Pdot}
\end{eqnarray}
in which $\Gamma$ is the mechanical loss rate, and $\xi\td$ is the thermal noise operator\footnote{The `quantum optics approximation' which distributes loss across both mechanical quadratures is not valid over short timesteps \cite{BowenMilburn}.}. The bath temperature determines the equilibrium phonon occupancy $\n_{M}$, which we will take to be much larger than one, such that $\expect{\xi\td\xi\Br{t^{\prime}} + \xi\Br{t^{\prime}}\xi\td} \approx 2\Br{2\n_{M}+1}\delta\Br{t-t^{\prime}}$.

The solution to equations~\eqref{Eqn:Xdot} and \eqref{Eqn:Pdot} is of the form
\[
	\bm{X} \rightarrow R_{M}^{\mathrm{loss}}\Br{\sigma\omega_{M}t}\bm{X} + \bm{F}_{M}\Br{t},
\]
where $R_{M}^{\mathrm{loss}}$ is a lossy rotation matrix, and $\bm{F}_{M}\td$ describes the effect of the thermal noise. The factor $\sigma = \sqrt{1-\Gamma^{2}/4\omega_{M}^{2}}$, which is approximately unity for high-$Q$ oscillators, describes the damping-induced retardation of the oscillator's period. Further details are given in \ref{App:Brownian}.

We must also consider optical loss. For simplicity, we will treat the case in which this is dominated by attenuation introduced in the delay line (\cf{} Fig.~\ref{Fig:Scheme},~panel \textit{b})), as would be the case for low-frequency mechanical oscillators requiring longer delay times. Under these circumstances we model optical loss as a beamsplitter interaction between the optical mode and a thermal mode with occupancy of $\n_{L}$. The beamsplitter's reflectivity is characterised by $\epsilon \in \Sq{0,1}$. Both quadratures are attenuated by a factor of $\sqrt{1-\epsilon}$ (represented by the matrix $L_{L}$), and each acquires an amount of thermal noise ($\bm{F}_{L}$) with variance $\epsilon\Br{2\n_{L}+1}$. Estimates can be found in \ref{App:OpticalLoss}.

With these two loss mechanisms in place, the full transformation between input and output operators is of the form $\bm{X}^{\prime} = \tilde{\mathcal{O}} \bm{X} + \bm{F}$, with
\begin{eqnarray*}
	\tilde{\mathcal{O}} & = & \mathcal{M}_{XX}\Br{\chi_{3}} R_{L}\Br{\theta-\frac{\pi}{2}} \mathcal{M}_{XX}\Br{\frac{-\lambda}{\cos\varphi}} R_{M}^{\mathrm{loss}}\Br{\varphi} \\
	& & {} \times L_{L}\Br{\epsilon} \mathcal{M}_{XX}\Br{\lambda} R_{L}\Br{\frac{\pi}{2}}\mathcal{M}_{XX}\Br{\chi_{1}},
\end{eqnarray*}
and the total incoming noise being
\[
	\bm{F} = \mathcal{M}_{XX}\Br{\chi_{3}} R_{L}\Br{\theta-\frac{\pi}{2}} \mathcal{M}_{XX}\Br{\frac{-\lambda}{\cos\varphi}}\Br{\bm{F}_{M} + \bm{F}_{L}}
\]
These results are obtained using equations~\eqref{Eqn:MPX} and \eqref{Eqn:Obar}, replacing $R_{M} \rightarrow R_{M}^{\mathrm{loss}}L_{L}$, and propagating the incoming noise through the final two pulses of the sequence.

\begin{figure}
\centering
\def\svgwidth{0.9\columnwidth}
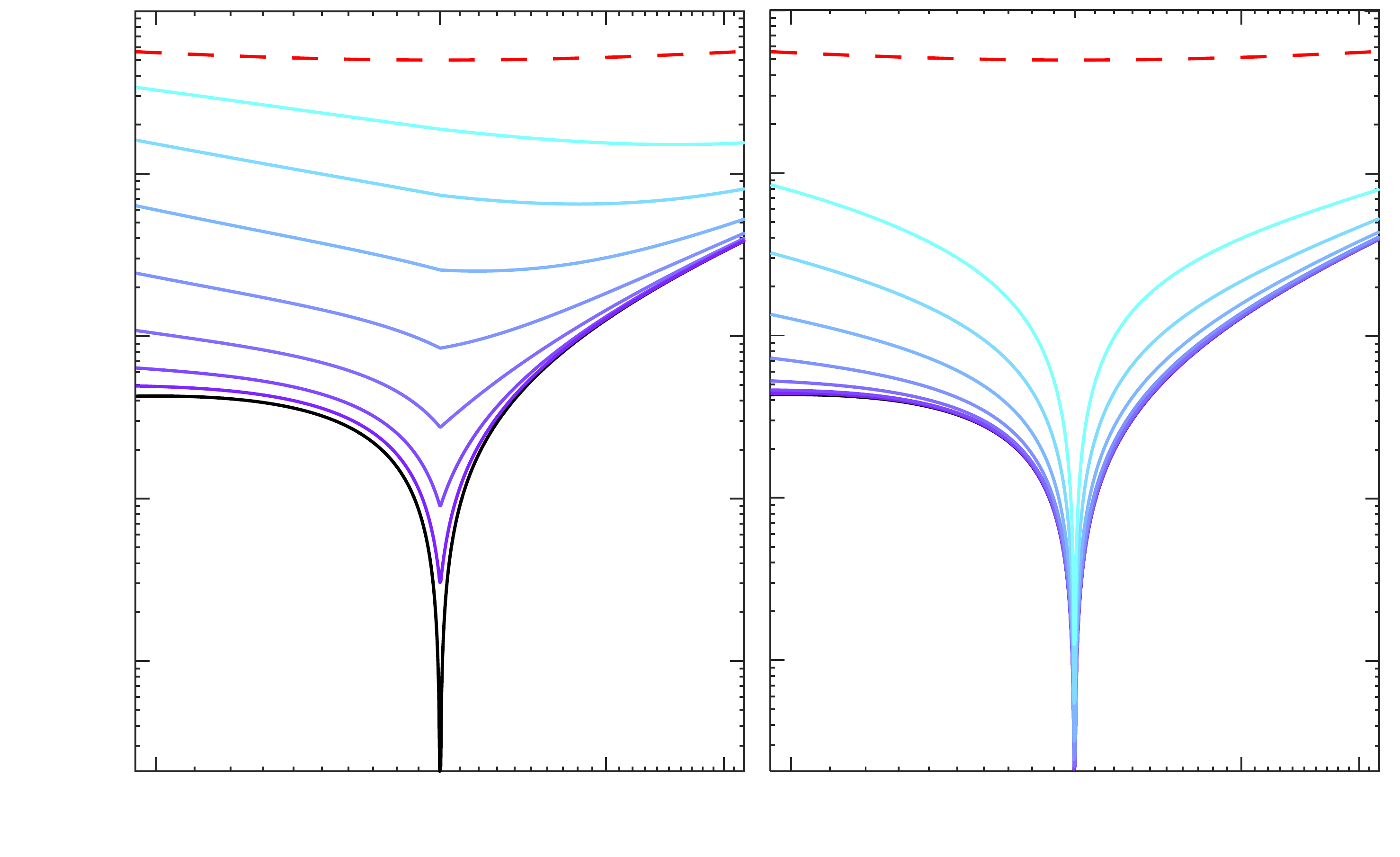
\caption{\label{Fig:Fidelity}
Minimum infidelity of mechanical squeezing as a function of the amount of squeezing ($\mu$) after one hundredth of a mechanical period. The mechanical oscillator begins in the ground state, and the optical ancilla has $3$~dB of squeezing on the appropriate quadrature. The ideal case is shown as black lines, and the classical limit (equation~\eqref{Eqn:ClassicalLimit}) is shown by red dashed lines.\\
\textit{a}) Degradation of infidelity due to mechanical damping. The darkest purple line corresponds to $Q = 10^{7}$; each line above that corresponds to a $\sqrt{10}$ reduction in $Q$, to a minimum of $10^{4}$. The thermal occupancy is set to $\n_{M} = 4\times 10^{4}$.\\
\textit{b}) Increase in infidelity due to optical loss. The darkest purple line shows $\epsilon = 10^{-5}$; every line above that increases $\epsilon$ by a factor of $\sqrt{10}$, to a maximum of $10^{-2}$. We have assumed that the loss introduces vacuum noise ($\n_{L} = 0$).
}
\end{figure}

Fig.~\ref{Fig:Fidelity} shows the effects of thermal noise (panel \textit{a}) and optical loss (panel \textit{b}) on the fidelity. The different shapes arise because thermal decoherence predominantly affects $P_{M}$, whereas the optical loss affects both optical quadratures equally. Note that the scheme still outperforms the classical protocol, even with a relatively modest mechanical $Q$ of $10^{4}$.

It is likely that numerically re-optimising the four pulse strengths could be used to compensate for some of the effects of decoherence, as demonstrated in previous proposals \eg{} \cite{Kupcik2015,Vostrosablin2017}. Other parameters such as the optical rotation angle $\theta$ and the ancilla squeezing angle $\phi$ could also be numerically adjusted. However, these optimisations do not serve to illuminate the proposed protocol, so we have not pursued them in this work.

\subsection{Photon number} \label{Sec:PhotonNumber}

\begin{figure}
\centering
\def\svgwidth{1.0\columnwidth}
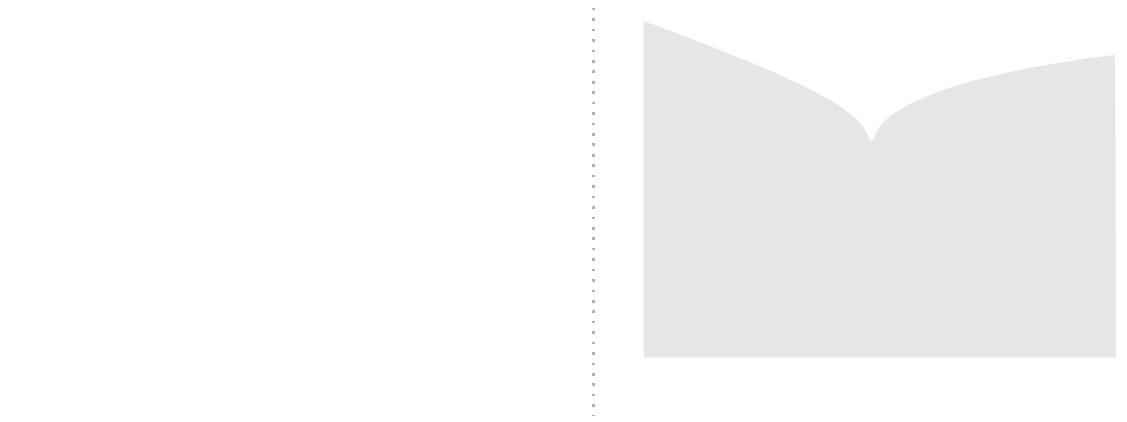
\caption{\label{Fig:Number}
\textit{a})~Total normalised photon number $\Lambda$ required to achieve a level of squeezing $\mu$. This approximate result holds for $\varphi \ll 1$, where the $\varphi$ dependence can be factorised. \\
\textit{b})~A specific realisation of optimum pulse strengths for various $\mu$ after a mechanical rotation of one hundredth of a cycle. The upper boundary of the shaded region corresponds to the value of $\chi_{1}$. Note the strong asymmetry of $\chi_{3}$. These values were calculated by numerically minimising the infidelity of the target state and the actual output (\cf{} equation~\eqref{Eqn:PureFidelity}). Note that this yields the same $\mod{\chi_{1}} = \mod{\lambda}$ result predicted by minimising $A$ at fixed $\mu$. We have chosen $\chi_{1} \geq 0$, but similar results can be achieved by flipping the signs of all pulses.
}
\end{figure}

In many experimental settings (\eg{} in cryostats) there is an upper limit to the number of photons that can be injected into the optomechanical cavity without causing unmanageable heating. This prompts us to consider the maximum (minimum) $\mu$ ($1/\mu$) that can be achieved subject to the restriction
\[
	\Lambda = \chi_{1}^{2}+\lambda^{2} + \lambda_{2}^{2} + \chi_{3}^{2} \leq \Lambda_{max},
\]
where $\Lambda$ can be interpreted as the total photon number (scaled by $64 g_{0}^{2}/\kappa^{2}$), and $\Lambda_{max}$ is some experimentally-imposed upper limit. Over small evolution angles $\varphi$ the required $\Lambda$ for a given amount of squeezing $\mu$ is well-approximated by the expression
\begin{equation}
	\Lambda\tan\varphi \approx \mod{1-\frac{1}{\mu}}\Br{3+\frac{1+\Br{1-\mu^{-1}}^{2}}{\mu^{2}}}.
	\label{Eqn:PhotonNumber}
\end{equation}
As can be seen in Fig.~\ref{Fig:Number} panel~\textit{a}), the number of photons required to achieve position squeezing ($\mu > 1$) is generally lower than for the equivalent level of momentum squeezing ($\mu \rightarrow \mu^{-1}$). For example, to achieve $3$~dB of position squeezing ($\expect{\left.X_{M}^{\prime}\right.^{2}} = 0.5$, corresponding to $\mu = \sqrt{2}$) over a hundredth of a mechanical period requires $N\sim 16\Br{\kappa/g_{0}}^{2}$, whereas the equivalent level of momentum squeezing ($\mu = 1/\sqrt{2}$) requires $N\sim 35 \Br{\kappa/g_{0}}^{2}$. This asymmetry is controlled by $\chi_{3}$.

To see why this is the case it is sufficient to consider the simple three-pulse interaction $\mathcal{O}$. The third pulse, being an $X$--$X$ QND interaction, cannot alter either $X_{M}$ or $X_{L}$, so any attenuation (amplification) of these operators must be complete by the end of pulse two. Pulse three maps this attenuated (amplified) amplitude noise onto $P_{M}$, where it can interfere with the noise contributed by pulse one. This can be seen in the lower left element of equation~\eqref{Eqn:Antidiagonal}, $\chi_{1} + \chi_{3}\Br{1+\chi_{1}\chi_{2}}$, where the first term corresponds to pulse one and the second to pulse three. The larger $1+\chi_{1}\chi_{2}$ is (the more amplification of $X_{M}$), the smaller $\mod{\chi_{3}}$ needs to be to achieve perfect cancellation. This leads to the asymmetry seen in Fig.~\ref{Fig:Number}.

\section{Performance: non-Gaussian inputs} \label{Sec:NonGaussian}
We have already established that our squeezer is both fast and nonclassical. This is attractive because it opens up the possibility of rapidly squeezing arbitrary mechanical input states, an ability unavailable to most other squeezing schemes because they are either incoherent \cite{Vitali2002,Ruskov2005,Jahne2009,Vanner2013}, slow \cite{Pirkkalainen2015a,Szorkovszky2011,Liao2011a,Agarwal2016,Wang2016,Mari2009,Nunnenkamp2010,Lu2015}, or greatly modify the final frequency of the mechanical oscillator \cite{Janszky1992}.

\begin{figure}
\centering
\def\svgwidth{0.9\columnwidth}
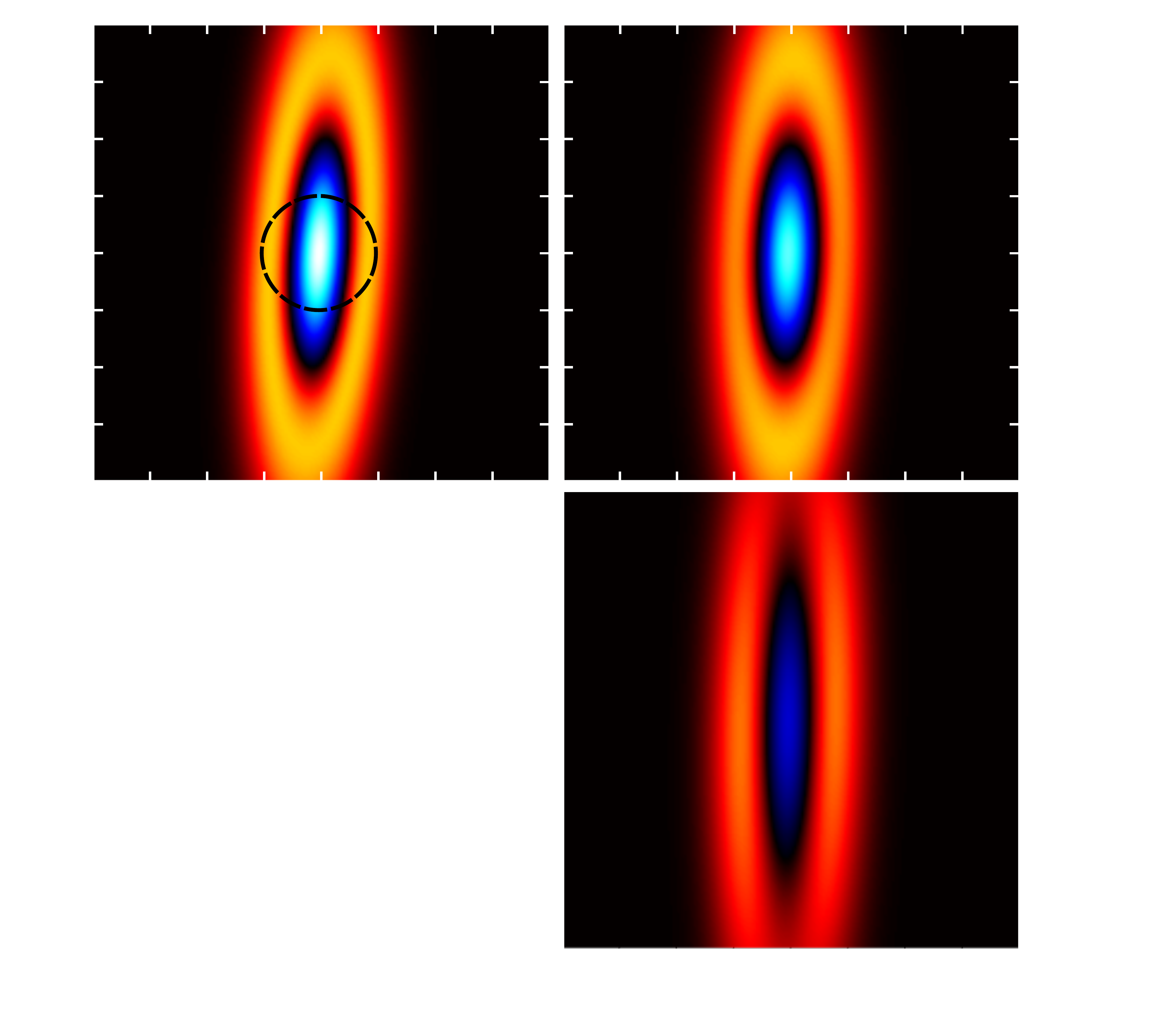
\caption{\label{Fig:Fock}
Demonstration of squeezing of non-Gaussian mechanical states. The initial $\ket{1}$ Fock state is squeezed with $\mu = 2$ over a hundredth of a mechanical period. 
Panel~\textit{a}) shows the target output state; panel~\textit{b}) shows the output of the ideal squeezer; and panels~\textit{c}) and \textit{d}) show the output under lossy squeezing with $\epsilon = 1\times 10^{-2}$ and $\epsilon = 5\times 10^{-2}$ respectively ($Q = 10^{5}$, $\n_{M}= 4\times 10^{4}$, $\n_{L} = 0$). The black dashed contours show the covariance ellipse of the ground state. Note that all cases show negativity at the origin, a clear indicator of nonclassicality.
}
\end{figure}

To illustrate the coherence of the scheme, in Fig.~\ref{Fig:Fock} we show the effect of our squeezer on a $\ket{1}$ Fock state. The single-phonon Fock state is an archetypical non-Gaussian state that exhibits strong Wigner negativity \cite{Kenfack2004}. This Wigner negativity is preserved by our scheme, albeit somewhat degraded by decoherence. Fig.~\ref{Fig:Fock} shows that a mechanical oscillator with $Q = 10^{5}$ and $\n_{M} = 4\times 10^{4}$, and an optical loss of on the order of one percent, can retain much of its Wigner negativity during one of these squeezing operations. An incoherent squeezer---such as a measurement-based squeezer \eg{} \cite{Vitali2002}---by contrast, would totally destroy the negativity.

These calculations were performed as detailed in our previous works \cite{Bennett2016}.

\section{Application: impulse detection} \label{Sec:Impulse}

Let us consider a simple protocol in which our rapid squeezer can improve the detection of an impulsive force of unknown amplitude arriving at a known time, $t = 0$.

Suppose the oscillator is initialised in a low-occupancy thermal state with an average of $\n_{in}$ phonons. The impulse arrives and displaces the momentum by $D$. The state is then allowed to evolve for a quarter of a cycle before $X$ is read out using a pulsed optomechanical interaction. If the measurement is capable of resolving the oscillator's initial momentum fluctuations (\ie{} not measurement noise limited) then we expect to be able to confidently resolve a minimum displacement $D$ equal to
\[
	D_{min} \sim \sqrt{2\n_{in}+1}.
\]
If we squeeze the momentum noise ($\mu < 1$) immediately before the displacement arrives we expect to improve this to
\[
	D_{min}\Br{\mu} \sim \mu\sqrt{2\n_{in}+1}.
\]

A more sophisticated expression can be obtained by incorporating damping into the $\pi/4$ rotation, treating the squeezer as ideal, and adding the shot noise from the optomechanical readout (defined by the strength $\chi_{ro}$). Doing so yields (in the high-$Q$, low $V_{sq}$ limit)

\begin{minipage}[H]{1 \columnwidth}
\begin{eqnarray}
	D_{min}\Br{\mu} & \approx & \sqrt{\mu^{2} \bar{N}_{in} + \chi_{ro}^{-2} + \frac{1-\mu}{\mu} V_{sq}\tan\varphi} \times \left[ 1+ \frac{\Gamma}{8\omega_{M}}\times  \right. \label{Eqn:Detection} \\
	& & \left. \frac{\pi\chi_{ro}^{-2} + \Br{\pi-2}\bar{N}_{M}+4\Br{1-\mu}\Br{\bar{N}_{in}\mu-\frac{V_{sq}}{\mu}}\tan\varphi}{\mu^{2} \bar{N}_{in} + \chi_{ro}^{-2} + \frac{1-\mu}{\mu}V_{sq}\tan\varphi}\right], \nonumber
\end{eqnarray}
\end{minipage}

where $\bar{N}_{in} = 2\n_{in}+1$ and $\bar{N}_{M} = 2\n_{M}+1$. This reduces to our previous estimate when the measurement is dominated by the noise of the mechanical input state.

\begin{figure}
\centering
\def\svgwidth{1\columnwidth}
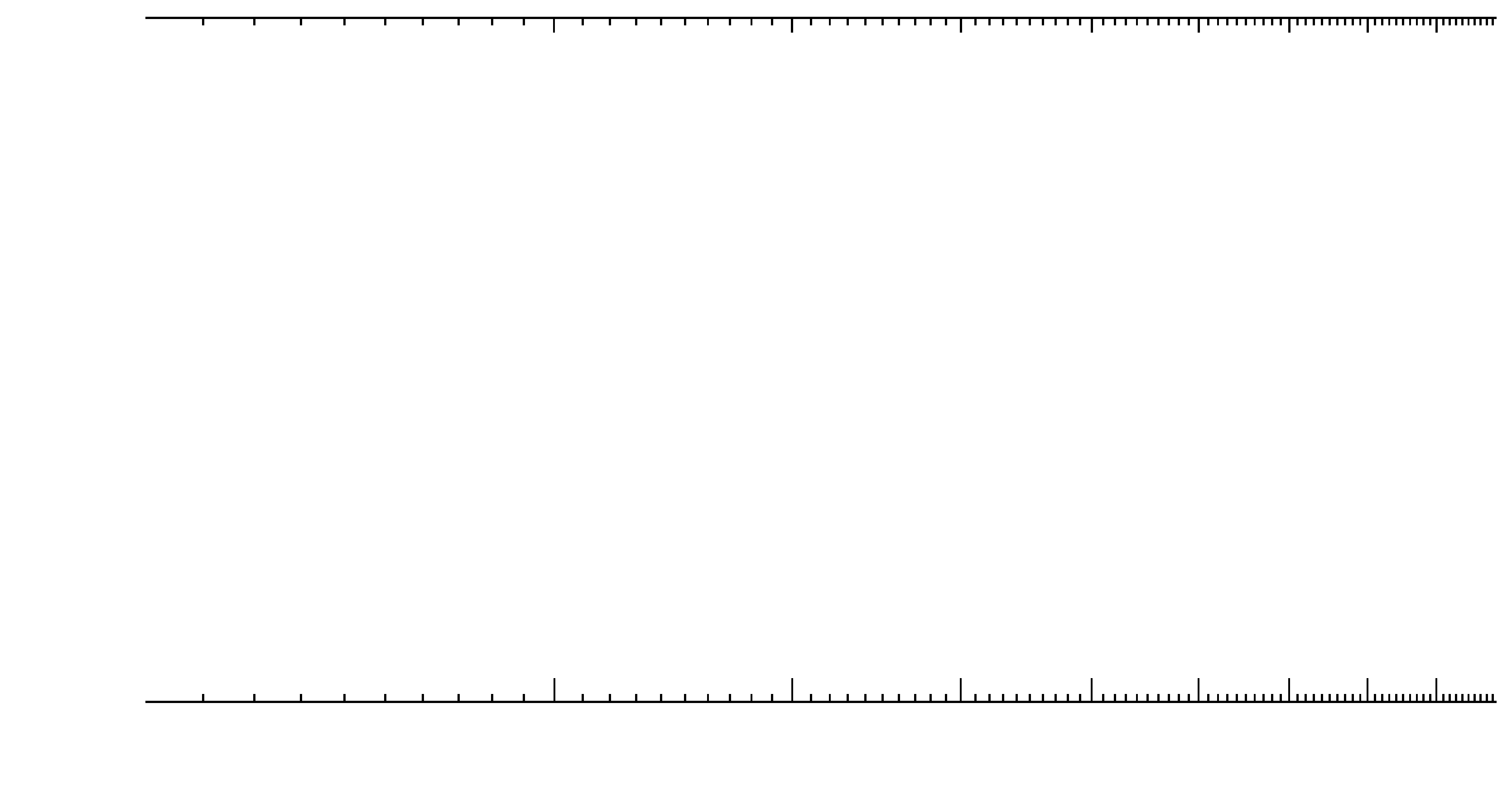
\caption{\label{Fig:Detection}
Minimum detectable impulse $D_{min}$ as a function of pre-squeezing $\mu$. It is clear that equation~\eqref{Eqn:Detection} (dashed lines) is a good approximation to the true result (solid lines) up to approximately $6$~dB of squeezing, particularly for higher initial occupancies. The \naive{} $D_{min} = \mu \sqrt{2\n_{in}+1}$ prediction fails for large amounts of squeezing because it does not account for imperfections in the squeezer. For all of these calculations the mechanical state is initially thermal, with occupancy $\n_{in}$; the optical ancilla for the squeezer has $3$~dB of squeezing ($V_{sq} = 0.5$). The final optomechanical readout step has $\chi_{ro} = 3$ and uses a coherent pulse. Other parameters are as given in Fig.~\ref{Fig:Fock}, panel~\textit{d}).
}
\end{figure}

A comparison of equation~\eqref{Eqn:Detection} with the full calculation (including thermal and optical decoherence during the squeezer) is given in Fig.~\ref{Fig:Detection}. As is evident, decreasing the input occupancy $\n_{in}$ and squeezing ($\mu < 1$) both improve the minimum detectable momentum kick. Shot noise and imperfections in the squeezer do degrade the sensitivity, but this effect appears manageable out to approximately $6$~dB of squeezing.

Surprisingly, to our knowledge, there has been little discussion of a scheme such as this for impulsive force sensing in the optomechanical literature, despite its conceptual simplicity. Khosla \etal{} \cite{Khosla2017} explored a closely-related protocol that leverages \textit{conditional} squeezing, and L\"{u} \etal{} \cite{Lu2015} briefly discussed the benefits of mechanical squeezing derived from a Duffing nonlinearity for impulsive force sensing. Other studies have focussed on using feedback to enhance impulsive force sensing \eg{} \cite{Vitali2001}.

\FloatBarrier

\section{Application: preservation of \Schrod{} cat states} \label{Sec:Cats}

Another potential application of a fast, coherent squeezer is the preservation of fragile quantum states, such as \Schrod{} cats \cite{Jeannic2018}. Cats formed by the superposition of coherent states exhibit strong Wigner negativity and are an important resource for continuous variable quantum communication \cite{Gilchrist2004}. The effect of squeezing on optical cat states has been considered theoretically by Filip \cite{Filip2001} and Serafini \etal{} \cite{Serafini2004}, and demonstrated experimentally by Le Jeannic \etal{} \cite{Jeannic2018}. Here we will consider storage of a cat state on a mechanical oscillator.

\begin{figure}
\centering
\def\svgwidth{1.0\columnwidth}
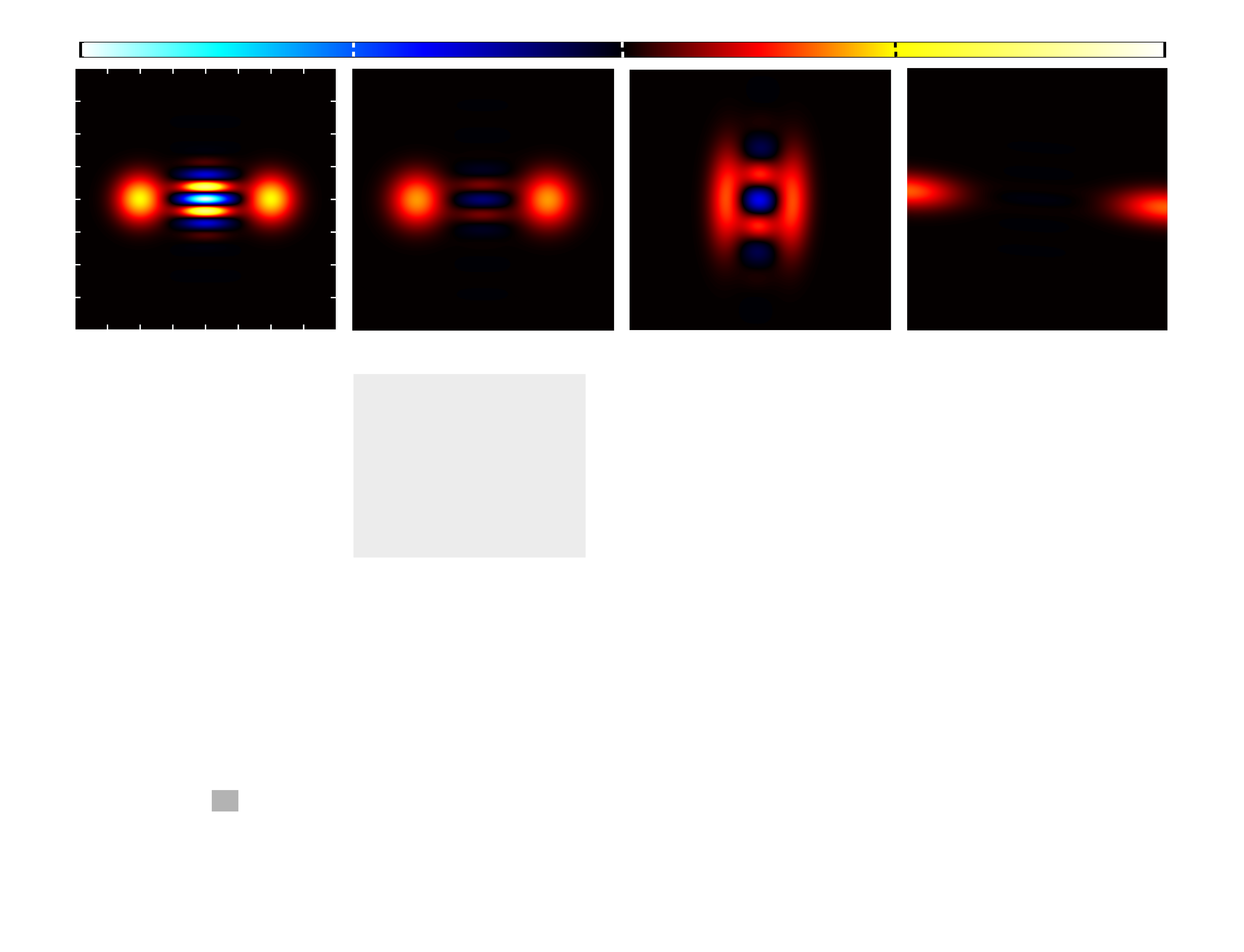
\caption{\label{Fig:Cats}
Slowing the decay of Wigner negativity of cat states using squeezing.\\
Panels \textit{a})--\textit{d}) show Wigner functions plotted on a $16\times 16$ grid centred on the origin of phase space.\\
\textit{a})~A pure, odd cat state with $\alpha = 2$, showing distinctive interference fringes along the vertical axis ($P_{M}$).\\
\textit{b})~The cat after five mechanical cycles with $\Gamma/\omega_{M} = 10^{7}$ and $\n_{M} = 4\times 10^{4}$; note the reduction in negativity, which occurs much more rapidly than the energy decay rate $\Gamma$.\\
\textit{c})~Pre-squeezing the cat can slow the loss of negativity. This state was momentum squeezed ($\mu = 2$, $\varphi = \pi/50$, $\epsilon = 10^{-3}$, $\n_{L}=0$) before undergoing five periods of damped evolution under the same conditions as above.\\
\textit{d})~Pre-squeezing with the wrong phase can make the state more susceptible to decoherence.\\
\textit{e}) The decay of negativity of $\alpha = 2$ odd cat states with and without pre-squeezing. An initial position squeezing step improves the negativity at long times, despite imperfections at short times. Momentum squeezing is everywhere detrimental. The inset shows a zoom-in on the grey square, revealing that the decay is periodically modulated by the asymmetry of damping on $X_{M}$ and $P_{M}$.\\
\textit{f})~A comparison of the negativity `half-life' with squeezing of different magnitudes for $\alpha = 1$ and $2$. Position squeezing $\mu>1$ is beneficial until the optimum $\mu_{opt}$ is achieved (see equation~\eqref{Eqn:MuOpt}). Red dots show the values obtained with no squeezing. Note that the improvement for $\alpha = 2$ is larger (in both absolute and relative terms) than for $\alpha = 1$.
}
\end{figure}

Cat states come in two flavours, odd ($-$) and even ($+$), defined by $\ket{\psi\sb{\pm}} \propto \ket{\alpha}\pm\ket{-\alpha}$. Coherence between the $\ket{\alpha}$ and $\ket{-\alpha}$ contributions forms an interference pattern in phase space. As can be seen in Fig.~\ref{Fig:Cats}, panel \textit{a}), the fringes are narrow in the direction orthogonal to $\alpha$ (here $\alpha = 2$). The effect of incoming thermal noise is essentially to convolve these fringes with a Gaussian kernel \cite{Bennett2016}, causing them to rapidly blend into one another and vanish (Fig.~\ref{Fig:Cats}, panel~\textit{b})). The kernel is approximately symmetric except at very short evolution times. We are therefore lead to expect that making a cat's fringes more symmetrical with squeezing can assist in preserving its nonclassical properties.

Potential outcomes are seen in Fig.~\ref{Fig:Cats}, panels~\textit{c}) and \textit{d}). As expected, symmetrising the fringes does preserve the Wigner negativity for longer; conversely, making the fringes more narrow makes the state more fragile to loss.

To quantify this effect we will restrict ourselves to considering odd cat states, which are negative at the origin of phase space. Our nonclassicality metric will be the normalised depth of this negativity, \viz{}
\begin{equation}
	\eta = \mathrm{max}\Cu{-2\pi W\Br{0},0},
	\label{Eqn:Negativity}
\end{equation}
where $W\Br{0}$ is the Wigner function evaluated at the origin. Thus $\eta = 1$ indicates that the cat has retained all of its negativity, whilst $\eta = 0$ indicates a non-negative Wigner function. The time evolution of $\eta$ is shown in Fig.~\ref{Fig:Cats}, panel~\textit{e}) for three cases: unsqueezed, position squeezed, and momentum squeezed. As predicted, position squeezing of the cat can symmetrise its fringes and improve its robustness to decoherence over multiple mechanical periods. Conversely, momentum squeezing makes the fringes narrower and degrades performance significantly.

Let us define $\tau$ as the time over which $\eta$ drops to $1/2$; this is analogous to a half-life, but cannot strictly be interpreted as such because the decay of the negativity is not exponential. Fig.~\ref{Fig:Cats}, panel~\textit{f}) shows $\omega_{M}\tau/2\pi$ as a function of $\mu$ for two different odd cats, $\alpha = 1$ and $\alpha = 2$. It is clear that an optimum amount of squeezing, $\mu_{opt}$, exists. This may be explained by considering the shape of the central negative fringe, which is well-approximated by the ellipse
\[
	\frac{\mu^{2}x^{2}}{2}\Br{\frac{4\alpha^{2}}{\e^{2\alpha^{2}}-1}+1} + \frac{p^{2}}{2\mu^{2}}\Br{\frac{4\alpha^{2}}{1-\e^{-2\alpha^{2}}}+1} = 1.
\]
This can be derived by expanding the Wigner function around the origin $x=p=0$. We can see that increasing $\mu$ tends to shorten the width along $x$---initially the longest direction---and lengthen it along $p$. Solving for the $\mu$ that yields a symmetric fringe (equal semi-major and semi-minor radii) yields
\begin{eqnarray}
	\mu_{opt} = \Sq{\frac{\e^{2 \alpha^{2}}\Br{1+4\alpha^{2}}-1}{\e^{2\alpha^{2}}+4\alpha^{2}-1}}^{1/4}.
	\label{Eqn:MuOpt}
\end{eqnarray}
This is a good approximation to the true behaviour seen in the figure; for $\alpha = 1$, $\mu_{opt} = 1.36$, whilst for $\alpha = 2$, $\mu_{opt} = 2.03$.

The other notable feature of Fig.~\ref{Fig:Cats}, panel~\textit{f}), is the periodic steps in $\tau$ as a function of $\mu$. These are in fact caused by the momentum-dependent damping of the oscillator (as in equations~\eqref{Eqn:Xdot} and \eqref{Eqn:Pdot}); the state decoheres more rapidly when its interference fringes are along $P_{M}$ than along $X_{M}$, causing a modulation of the negativity's decay rate at twice the mechanical resonance frequency. This effect can be seen in the inset to Fig.~\ref{Fig:Cats}, panel~\textit{e}), as well as panel~\textit{f}). 

\section{Conclusion} \label{Sec:Conclusion}

We have proposed a fast and coherent squeezer based on four optomechanical pulses. We predict the possibility of squeezing of arbitrary mechanical inputs, including non-Gaussian states, with reasonable fidelity and retention of negativity even in the presence of decoherence. Our protocol is suitable for use with solid-state oscillators of fixed frequency, and may be adapted for use with other physical systems that permit rapid QND interactions. This work may find application in quantum information technologies, where it could be used to improve the storage of phononic \Schrod{} cat states \ala{} the work of Jeannic \etal{} \cite{Jeannic2018}.

\section*{Acknowledgments} \label{Sec:Acknowledgments}

The authors thank Kiran Khosla and Yauhen Sachkou for discussions concerning pulsed optomechanics, and Christiaan Bekker and Stefan Forstner for critique of figures. This work was funded by the Australian Research Council (ARC), CE110001013. JSB acknowledges support from an Australian Government Research Training Program Scholarship. WPB is an ARC Future Fellow (FT140100650).

\section{References} \label{Sec:Ref}

\providecommand{\newblock}{}

\appendix

\section{Effect of multiple mechanical modes} \label{App:AdditionalMechModes}

This scheme---and others based on sub-period pulses---require that the pulse bandwidth $\Omega_{p}$ satisfy $\omega_{M} \ll \Omega_{p} \ll \kappa \ll \omega_{cav}$. The first inequality ensures that the pulse is much shorter than the mechanical period, the second prevents it from being distorted by the cavity's response function, and the third ensures that the optical $Q$ is high. In this situation the pulse couples to a single cavity mode, with other optical modes being highly off-resonant. However, the same cannot be said for nearby \textit{mechanical} modes.

The mechanical spectrum of most solid-state resonators is likely to have many modes residing within $\Omega_{p}$ of the mode of interest. This problem has been acknowledged in previous literature (specifically \cite{Vanner2011,Vanner2013}), but most proposals do not explicitly account for these undesired couplings.

If we suppose that there are $j = 1,2,..., j_{max}$ mechanical modes that each couple to the one optical cavity then Eqns~\eqref{Eqn:QNDX} and \eqref{Eqn:QNDP} still hold, but with the replacements
\begin{eqnarray*}
	X_{M} & \rightarrow & \Br{\sum\sb{j=1}^{j_{max}} g\sb{j}}^{-1} \sum\sb{j=1}^{j_{max}} g\sb{j}X\sb{j}, \\
	P_{M} & \rightarrow & \sum\sb{j=1}^{j_{max}} P\sb{j}, \\
	\chi & \rightarrow & -\frac{8\sqrt{N}}{\kappa} \sum\sb{j=1}^{j_{max}} g\sb{j}.
\end{eqnarray*}
These are canonically conjugate operators obeying $\comm{X_{M}}{P_{M}}=2\I$. Thus the major difference between our single-mode treatment and the multi-mode case given here is the free evolution of the collective mode $X_{M}$, which is no longer harmonic.

A full multi-mode treatment of this squeezing scheme is beyond the scope of this work. However, we can obtain some simple estimates of its effects by adding a single auxiliary mechanical mode with frequency $\omega_{2}$ and coupling rate $g_{2}$. For simplicity, we will set $\Gamma = 0$, $\epsilon = 1$, $\varphi = \pi/50$, and $\omega_{2} = 2\omega_{M}$. If we begin with all three modes in vacuum states and aim for $\mu = \sqrt{2}$ ($3$~dB of position squeezing), then the infidelities ($I$) behave as follows; with $g_{2} = g_{1}$, $I = 0.70$; $g_{2} = g_{1}/2$, $I = 0.19$; $g_{2} = g_{1}/5$, $I = 0.031$; and $g_{2} = g_{1}/10$, $I = 0.021$. Compare these to the single-mechanical-mode fidelity of $I = 0.018$. It is clear that reducing the coupling $g_{2} \rightarrow 0$ reduces the infidelity towards the single-mode result, and this process occurs rather rapidly with decreasing $g_{2}$. This indicates that the problem is likely to be manageable even for many modes, provided that the sum of the unwanted couplings can be reduced sufficiently below the desired $g_{1}$ (\ie{} $\sum\sb{j\neq 1} g\sb{j} < g_{1}$).

As already noted by Vanner \etal{} \cite{Vanner2011,Vanner2013}, there are two main routes to addressing stray couplings. Firstly, one can engineer the mechanical spectrum by controlling the geometry, materials, and processing of the oscillator. An extreme example of this is using trapped ions, atoms, or dielectrics as mechanical elements; these systems have clean mode spectra that can be tuned \textit{in situ}. Secondly, one can reduce the $g\sb{j}$ of unwanted contributions by exploiting symmetries to reduce mode overlaps, or by increasing the effective mass of undesired modes.

A comprehensive treatment of this problem is currently lacking, and would be a valuable tool for designing experimental realisations of these protocols.

\section{Thermal loss} \label{App:Brownian}

The solution to equations~\eqref{Eqn:Xdot} and \eqref{Eqn:Pdot} is
\[
	\bm{X} \rightarrow R_{M}^{\mathrm{loss}}\Br{\sigma\omega_{M}t}\bm{X} + \bm{F}_{M}\Br{t}.
\]
with
\[
	R_{M}^{\mathrm{loss}} = \Br{\begin{array}{cc}
	\begin{array}{cc}
	\e^{-\frac{\Gamma t}{2}}\Br{\begin{array}{c}
	\cos\Br{\sigma\omega_{M}t} + \\
	\frac{\Gamma}{2 \sigma \omega_{M}}\sin\Br{\sigma\omega_{M}t} \end{array}} & +\frac{\e^{-\frac{\Gamma t}{2}}}{\sigma}\sin\Br{\sigma\omega_{M}t} \\
-\frac{\e^{-\frac{\Gamma t}{2}}}{\sigma}\sin\Br{\sigma\omega_{M}t} & \e^{-\frac{\Gamma t}{2}}\Br{\begin{array}{c}
\cos\Br{\sigma\omega_{M}t} - \\
\frac{\Gamma}{2 \sigma \omega_{M}}\sin\Br{\sigma\omega_{M}t} \end{array}}
	\end{array}
	& \bm{0} \\
	\bm{0} & \identity
	\end{array}}.
\]
Here $\bm{0}$ is a $2\times 2$ matrix of zeros, and $\identity$ is a $2\times 2$ identity matrix. The covariance of the thermal noise $\bm{F}_{M}\td$ is
\begin{eqnarray}
	V_{FF} & = & \intfin{t^{\prime}}{0}{t} \intfin{t^{\prime\prime}}{0}{t} 2\Gamma\Br{2 \n_{M}+1}\delta\Br{t-t^{\prime}} M\Br{t-t^{\prime}} \Br{\begin{array}{cc} 0 & 0 \\ 0 & 1 \end{array}}\trans{M}\Br{t-t^{\prime\prime}} \nonumber \\
	& = & \Br{\begin{array}{cc}
		V_{FF}^{1,1} & V_{FF}^{1,1} \\ V_{FF}^{1,2} & V_{FF}^{2,2} \end{array}}
	\label{Eqn:ThermalEvolutionVFF}
\end{eqnarray}
with
\begin{eqnarray*}
		V_{FF}^{1,1} & = & \Br{2\n_{M}+1}\Sq{1+\frac{\e^{-\Gamma  t}}{\sigma^{2}}\Br{\frac{\Gamma^{2}\cos\Br{2 \sigma \omega_{M} t}-2\Gamma\sigma\omega_{M}\sin\Br{2\sigma\omega_{M} t}}{4\omega_{M}^{2}}-1}}, \\
		V_{FF}^{2,2} & = &  \Br{2\n_{M}+1}\Sq{1+\frac{\e^{-\Gamma  t}}{\sigma^{2}} \Br{\frac{\Gamma^{2}\cos\Br{2\sigma \omega_{M} t}+2\Gamma\sigma\omega_{M}\sin\Br{2\sigma\omega_{M} t}}{4 \omega_{M}^{2}}-1}}, \\
		V_{FF}^{1,2} & = & \Br{2\n_{M}+1}\Sq{\frac{\Gamma \e^{-\Gamma t}}{\sigma^{2} \omega_{M}} \sin^{2}\Br{\sigma\omega_{M}  t}}.
\end{eqnarray*}
Over short timescales the dominant contribution is $V_{FF}^{2,2}$

\section{Optical loss} \label{App:OpticalLoss}

There are three primary locations that optical loss may be introduced into our protocol: the optical cavity, the partially reflective mirrors, and the optical delay line. We expect that the effects of these different loss channels to be qualitatively similar; the results presented in this paper are quantitatively correct for loss in the delay line.

Loss from the cavity can be made negligible by strongly overcoupling to the input--output channel (\ie{} having a high escape efficiency). Partially reflective mirrors (used to perform rotations of the optical noise) are required at four points during the protocol. In principle, the noise associated with these can be made arbitrarily small so long as one has access to sufficiently powerful displacing pulses. We will operate in this limit. That leaves loss from the delay line.

Supposing that the delay line is constructed from a length of optical fibre, one might expect a loss of approximately $0.4$~dB/km. Suppose that $\omega_{M} \sim 10^{6}$ and the time delay is one hundredth of a mechanical period ($t = \varphi/\omega_{M} \sim 6\times 10^{-8}$~s). This requires a fibre of length $\ell \sim 12$~m, for a total loss of $\sim 0.005$~dB. This corresponds to a beamsplitter ratio of $\epsilon \approx 10^{-3}$.

\end{document}